\documentclass[12pt]{article}

\def\be{\begin{equation}}
\def\ee{\end{equation}}

\newtheorem{theorem}{Theorem}
\newtheorem{lemma}[theorem]{Lemma}

\newtheorem{definition}[theorem]{Definition}

\begin{document}

\title{Replica symmetry breaking \\ related to a general ultrametric space II: \\
RSB solutions and the $n\to0$ limit}

\author{A.Yu.Khrennikov\footnote{International Center for Mathematical
Modelling in Physics and Cognitive Sciences, University of
V\"axj\"o, S-35195, Sweden, e--mail:
Andrei.Khrennikov@msi.vxu.se}, S.V.Kozyrev\footnote{Steklov
Mathematical Institute, Moscow, Russia, e--mail:
kozyrev@mi.ras.ru}}

\maketitle

\begin{abstract}
Replica symmetry breaking solutions for the new replica anzats,
related to general ultrametric spaces, are investigated. A variant
of analysis on trees is developed and applied to the computation
of the $n\to0$ limit in the new replica anzats.
\end{abstract}

\section{Introduction}

In the present paper, continuing the line of research of
\cite{RSBI}, we introduce the variant of the $n\to0$ limit of
replica approach, suitable for the new family of replica matrices,
introduced in  \cite{RSBI}, and investigate the replica symmetry
breaking (RSB) solutions.

One of the most interesting phenomena of replica theory of spin
glasses and other disordered systems is the property of
ultrametricity of the replica space \cite{nature}. In papers
\cite{ABK} and \cite{PaSu} it was shown that, in important
particular case, this ultrametric structure of replica space can
be described with the help of $p$--adic analysis. For general
introduction to the replica method see \cite{MPV}.

In papers \cite{ACHA}--\cite{nextIzv} a very general family of
ultrametric spaces was constructed, and a theory of ultrametric
pseudodifferential operators (or PDO) was developed. The mentioned
above results are related to the field of $p$--adic and
ultrametric mathematical physics. For the other developments in
this field see \cite{VVZ}--\cite{ABKO}.

In \cite{RSBI} the family of replica matrices of very general form
(see (\ref{Qmatrix}) below), related to ultrametric PDO of
\cite{ACHA}, \cite{Izv}, was proposed and some functionals of the
replica approach for these matrices were computed.

In the present paper, continuing this line of research, we
introduce the $n\to0$ limit procedure, suitable for the replica
anzats under investigation. We compute some functionals of replica
approach in the $n\to0$ limit. We show that computation of these
functionals is related to some kind of analysis on directed trees,
with the corresponding tree derivation and integration. We
introduce the mentioned tree derivation and integration and find
the corresponding tree Leibnitz rule, the tree Newton--Leibnitz
formula, and the other constructions of the analysis on directed
trees. In particular, the functionals of the replica method will
have the form of tree integrals, and computation of the
functionals will use the constructions of the analysis on directed
trees.

The first of the main results of the present paper is the
following. We show that in the framework of the RSB anzats of
\cite{RSBI} there exist at least two different families of replica
matrices, for which it is possible to perform the $n\to 0$
procedure. For the first family, when the matrix element of the
replica matrix defined as in \cite{RSBI} (see the Appendix for the
notations): \be\label{Qmatrix} Q_{IJ}=\sqrt{\mu(I)\mu(J)}T{(\,{\rm
sup}\,(I,J))},\qquad I,J\in {\cal S}_{\rm min} \ee with
$$
T{(J)}=F(\mu(J))
$$
(which is the generalization of the Parisi anzats), the
functionals of replica approach take the form (as for the Parisi
anzats) of the integral over the unit interval
$$
-\int_0^1F(x)dm(x)
$$
where $F$ is some function and $m$ is some measure on the interval
$[0,1]$.

For the second family of replica matrices (which are not analogous
to the considered before), matrix elements are defined by
(\ref{Qmatrix}) with $T{(J)}$ satisfying to equation
$$
\Delta \left[\mu(L)T{(L)}\right]=0
$$
(i.e. is a constant of the tree derivation $\Delta$). In this case
the corresponding replica functionals take the form
$$
-\lim_{K\to\infty}{1\over\mu(K)}\int_{K}\phi(x)d\mu(x)
$$
of the limit of the normed integrals over the increasing family of
disks in ultrametric space, where the $\phi$ is some nonnegative
generalized function on the ultrametric space. Therefore, for the
different examples of replica matrices the $n\to0$ limits of the
functionals may take the form of integrals over real as well as
over ultrametric parameters.

Then, we find the replica symmetry breaking equation, which is
obtained by minimization of the free energy in the frameworks of
the investigated RSB anzats (\ref{Qmatrix}). We find two solutions
of the replica symmetry breaking equation. The first is the
constant solution, for which $T{(L)}=T$ for all $L$, and the
constant $T$ is determined by parameters of the model.

The second is the generalization of the Parisi RSB solution onto
the case of general ultrametric space, which is given by
$$
T{(L)}= {\rm min}\,\left[{a_3\over 4a_4}\rho(\mu({L})),T\right]
$$
where $a_3$, $a_4$ are the coefficients. This is the second of the
main results of the present paper.

The organization of the present paper is as follows.

In Section 2 we introduce the analysis on directed trees.

In Section 3 we introduce a variant of the $n\to0$ limit, suitable
for the replica symmetry breaking anzats under consideration.

In Section 4, using the tree analysis of Section 2, we compute the
functionals of replica matrices and their $n\to0$ limits.

In Section 5 we find the replica symmetry breaking equation.

In Section 6 we investigate the constant solution for this
equation, together with the $n\to0$ limit for this solution.

In Section 7 we find the solution with broken replica symmetry,
which is the analogue of the Parisi RSB solution in the case of
general ultrametric spaces.

In Section 8 (the Appendix) we put some material on trees and
ultrametric spaces.

\section{Analysis on trees}

In the present section we discuss the analysis on directed trees.
We define tree derivation and integration over the subtrees of the
regular type.

\begin{definition}{\sl For the function $F(J)$ on the directed tree ${\cal
T}$ the function \be\label{Delta} \Delta F(J)=
F(J)-\sum_{j=0,\dots,p_J-1:J_j<J,|JJ_j|=1}F(J_j) \ee we call the
tree derivative.

The tree integral over the subtree of the regular type ${\cal
S}\subset {\cal T}$ we define as
$$
\int_{\cal S} F=\sum_{J\in {\cal S}\backslash {\cal S}_{\rm
min}}F(J)
$$
}
\end{definition}
Here ${\cal S}_{\rm min}$ is the set of minimal elements in ${\cal
S}$, distance $|IJ|$ between vertices of the tree is the number of
edges in the path connecting $I$ and $J$, $p_J$ is the branching
index of $J$. Summation in (\ref{Delta}) runs over maximal
vertices which less than $J$.

In the following in the notation for the tree derivative for
simplicity instead of (\ref{Delta}) we use the simplified notation
$$
\Delta F(J)= F(J)-\sum_{j}F(J_j)
$$
Examples of the tree derivatives:
$$
\Delta \mu^n(J)=\mu^{n}(J)\left(1-p_J^{1-n}\right),
$$
in particular,
$$
\Delta \mu(J)=0;
$$
$$
\Delta 1=1-p_J.
$$

The next lemma relates the analysis on the directed tree and the
analysis on the absolute of the tree. This shows, that generalized
functions at the absolute can be considered as the constants of
the tree derivation.

\begin{lemma}
{\sl The space of solutions of the equation
$$
\Delta F(J)=0
$$
is isomorphic, as a linear space, to the space of generalized
functions at the absolute $X({\cal T})$, with the isomorphism
defined by the formula
$$
\phi_F(\chi_J)=F(J)
$$
Here $\phi_F$ is the generalized function at the absolute,
corresponding to the function $F$ at the tree, and $\chi_J$ is the
characteristic function of the disk $J$. }
\end{lemma}

\noindent{\it Proof}\qquad The proof is by the remark that a
generalized function on the ultrametric space $X$ is defined
unambiguously by its values (as of the functional) on the
characteristic functions of disks.

Characteristic functions of disks are not linearly independent,
but are related as follows
$$
\chi_{J}=\sum_{j=0,\dots,p_J-1:J_j<J,|JJ_j|=1}\chi_{J_j}
$$
By linearity of generalized functions, this implies the following
conditions of the values of generalized functions:
$$
\phi(\chi_J)=\sum_{j=0,\dots,p_J-1:J_j<J,|JJ_j|=1}\phi(\chi_{J_j})
$$
Choosing $F(J)=\phi(\chi_J)$, we get exactly
$$
\Delta F(J)=0
$$
Since no other restrictions on $F(J)$ are put, this proves the
lemma.

\bigskip

The following generalization of the above lemma is obvious.

\begin{lemma}\label{lemma2.3}
{\sl The space of solutions of the equation
$$
p_J^nF(J)-\sum_{j=0,\dots,p_J-1:J_j<J,|JJ_j|=1}F(J_j)=0
$$
is isomorphic, as a linear space, to the space of generalized
functions at the absolute $X({\cal T})$, with the isomorphism
defined by the formula
$$
\phi_F(\chi_J)=F(J)\mu^{n}({J})
$$
Here $\phi_F$ is the generalized function at the absolute,
corresponding to the function $F$ at the tree, and $\chi_J$ is the
characteristic function of the disk $J$. }
\end{lemma}

There exist several analogies between the introduced analysis on
trees and the analysis of functions of real argument.

There exists the following partial analogue of the Leibnitz rule
\be\label{LeibRule} \Delta
F(J)G(J)=F(J)G(J)-\sum_{j=0,\dots,p_J-1:J_j<J,|JJ_j|=1}F(J_j)G(J_j)=
$$
$$
=F(J)G(J)-\sum_{j}F(J_j)G(J)+\sum_{j}F(J_j)G(J)-\sum_{j}F(J_j)G(J_j)=
$$
$$
=[F(J)-\sum_{j}F(J_j)]G(J)+\sum_{j}F(J_j)[G(J)-G(J_j)] \ee

The next lemma gives the tree analogue of the Newton--Leibnitz
formula
$$
\int_a^b{df(x)\over dx}dx=f(b)-f(a)
$$
This lemma is of crucial importance for replica computations and
shows the importance of the notion of a subtree of the regular
type.

\begin{lemma}{\sl For the subtree ${\cal S}\subset {\cal T}$ of the regular type the following tree Newton--Leibnitz
formula is satisfied
$$
\int_{\cal S}\Delta F(J)=F(K)-\sum_{J\in
 {\cal S}_{\rm min}}F(J)
$$
The important examples of this formula are
$$
\int_{\cal S}\mu^2(J)\left(1-p_J^{-1}\right)=\mu^2(K)-\sum_{J\in
 {\cal S}_{\rm min}}\mu^2({J})
$$
$$
\int_{\cal S}(p_J-1)=-1+\sum_{J\in
 {\cal S}_{\rm min}}1
$$
}
\end{lemma}

In the next two formulas the tree derivative is taken with respect
to $J$.

\begin{lemma}{\sl
\be\label{higher} \Delta \sum_{L\in {\cal S}\backslash {\cal
S}_{\rm min}:L\le J}F(L)=F(J) \ee \be\label{lower} \Delta
\left[\mu(J)\sum_{L:J\le L\le
K}F(L)\right]=-\mu({J-1})\sum_{j=0,\dots,p_J-1:J_j<J,|JJ_j|=1}F(J_j)
\ee }
\end{lemma}

The formula (\ref{higher}) (respectively (\ref{lower})) is the
tree analogue of the derivative of the integral over the higher
(respectively the lower) limit.

\bigskip

The next lemma is the analogue of the following change of the
order of integration:
$$
\int_a^b
f(x)\left[\int_a^xg(y)dy\right]dx=\int_a^bg(y)\left[\int_y^bf(x)dx\right]dy
$$

\begin{lemma}\label{change}{\sl
$$
\sum_{L\in {\cal S}\backslash {\cal S}_{\rm min}} F(L)\sum_{B\in
{\cal S}\backslash {\cal S}_{\rm min} :B<L}G(B)=\sum_{L\in {\cal
S}\backslash {\cal S}_{\rm min}}G(L)\sum_{B:L<B\le K}F(B)
$$
}
\end{lemma}

The above sums are the analogues of integration over several
variables.

\section{The $n\to 0$ limit: definition}

The present and the next sections are written at the physical
level of rigor. In the present section we describe the
generalization of the $n\to 0$ limit of replica approach, relevant
to the introduced replica symmetry breaking (RSB) anzats.

Consider the map $\eta$, which acts on the measures $\mu(J)$ of
ultrametric disks according to the following rules:

1) Normalization: \be\label{norm42} \eta(\mu(R))=1 \ee where $R$
is the root of the tree, for which $\mu(R)=1$.

2) Monotonicity and infinitesimality: \be\label{mult}
\eta(\mu(J))>\eta(\mu(I)) \ee for $I>J$ and, moreover,
\be\label{mult1} d\eta(\mu(J))=\eta(\mu(J))-\eta(\mu({J+1}))
\qquad \hbox{is~ a~ positive~ infinitesimal~ value}\ee where $J+1$
is the smallest vertex larger than $J$.

3) Vanishing of the limit:
\be\label{dpdp}\lim_{I\to\infty}\eta(\mu(I))=0.\ee

\bigskip

We will perform computations with subtrees ${\cal S}\subset {\cal
T}$ of regular type. We will take $R\in {\cal S}_{\rm min}$ and
will claim, that $\eta(\mu(J))$ for $J\in {\cal S}_{\rm min}$
should be equal to 1 up to infinitesimal corrections which we will
neglect.

The rule (\ref{dpdp}) means that the limit $n\to 0$ is related to
the limit $I\to\infty$ in the directed tree. Thus our construction
indeed is a variant of the $n\to 0$ limit, since in our approach
$\mu(I)$ coincides with the dimension $n$ of the replica matrix
(when $I$ is the maximal vertex in the subtree ${\cal S}$ of the
regular type).

Condition (\ref{mult1}) implies that in the $n\to0$ limit for any
$J$ one has $p_J\to 1-\varepsilon$ for the infinitesimal
$\varepsilon$. Some variant of the analogous construction was
described in paper \cite{PaSu}, where, in the $p$--adic case, the
$n\to 0$ was discussed as the map $p\mapsto 1-\varepsilon$,
$\varepsilon\to 0$, which is the analogue of the formula
(\ref{mult1}).

Then, we introduce the $n\to 0$ limit in the RSB anzats under
consideration as the map $\rho$, which acts on the polynomials
over the variables, equal to the measures of the ultrametric disks
$\mu(J)$. This map is linear with respect to addition and
multiplication by numbers, and action on the monomials of $\mu(J)$
is defined as follows: \be\label{namonomy}
\rho\left(\mu^{k}(J)\right)=\mu(J)\eta^{k-1}\left(\mu(J)\right)
\ee

The formulas (\ref{norm42})--(\ref{dpdp}) are the direct analogues
of the definitions of the $n\to0$ limit for the Parisi anzats, and
the formula (\ref{namonomy}) is the new condition which was
trivial for the Parisi anzats, an becomes nontrivial in the case
under consideration.

\bigskip

\noindent{\bf Remark}\qquad The described procedure of the $n\to
0$ limit is ambiguous. In particular, transformations of the
polynomial over $\mu(J)$ and the $n\to 0$ limit does not commute.
Therefore there is an analogy between the quantization procedure
and the taking of the $n\to 0$ limit procedure: the both
constructions are ambiguously defined. Can the $n\to0$ limit be
connected with noncommutative probability, is not clear at the
present moment.

\section{The $n\to 0$ limit: examples}

Investigate the introduced $n\to0$ limit construction in some
important particular cases. Investigate the functional
\be\label{thefunc0} {1\over n}\sum_{ab}Q_{ab}= {1\over
\mu(K)}\sum_{J\in {\cal S}\backslash {\cal S}_{\rm
min}}T{(J)}\mu^2(J)\left(1-p_J^{-1}\right) \ee in the $n\to 0$
limit.

\bigskip

Consider the case, which is the direct generalization of the
Parisi anzats for the case of general ultrametric space. We have
the following lemma.

\begin{lemma}{\sl For the case, when the replica matrix is defined
by the function of measures of ultrametric disks \be\label{QP}
T{(J)}=F(\mu(J)) \ee the $n\to0$ limit of the functional
(\ref{thefunc0}), in the case when the function $F$ is continuous
in the interval $[0,1]$, takes the form \be\label{limthefunc2}
\lim_{n\to0}{1\over n}\sum_{ab}Q_{ab}=-\int_0^1 F(x)dm(x) \ee
where the measure $dm(x)$ on the interval $[0,1]$ is defined as
$$
\int_0^1 F(x)dm(x)=\lim_{K\to\infty}{1\over \mu(K)}\sum_{J\in
{\cal S}\backslash K}F(\rho(\mu(J)))\mu({J})d\rho(\mu({J}))
$$
 }
\end{lemma}

\noindent{\it Proof}\qquad Consider the functional
$$
{1\over \mu(K)}\sum_{J\in {\cal S}\backslash {\cal S}_{\rm
min}}F(\mu(J))\mu^{2}(J)\left(1-p_J^{-1}\right)= {1\over
\mu(K)}\sum_{J\in {\cal S}\backslash {\cal S}_{\rm
min}}F(\mu(J))\Delta\mu^{2}(J)
$$

By (\ref{LeibRule}) we have
$$
\rho(\Delta\mu^2(J))=-\sum_{j=0}^{p_J-1}\mu({J_j})d\rho(\mu({J_j}))
$$
Therefore
$$
\rho\left[{1\over \mu(K)}\sum_{J\in {\cal S}\backslash{\cal
S}_{\rm min}}T{(J)}\Delta \mu^2(J)\right]=-{1\over
\mu(K)}\sum_{J\in {\cal S}\backslash
K}\rho\left[T{(J+1)}\right]\mu({J})d\rho(\mu({J}))
$$
where $J+1$ is the unique minimal vertex, larger than $J$. For the
case $T{(J)}=F(\mu(J))$ this takes the form
$$
-{1\over \mu(K)}\sum_{J\in {\cal S}\backslash
K}F(\rho(\mu({J}))-d\rho(\mu({J})))\mu({J})d\rho(\mu({J}))
$$
which for continuous $F$ reduces to
$$
-{1\over \mu(K)}\sum_{J\in {\cal S}\backslash
K}F(\rho(\mu({J})))\mu({J})d\rho(\mu({J}))
$$
Application of the $K\to\infty$ limit proves the lemma.

\bigskip

Consider the new case, in which the functional under investigation
will be given by integration over the absolute of the tree.

Remind that, by lemma \ref{lemma2.3}, the space of solutions of
the system of equations
$$
\sum_{j=0}^{p_L-1}\left(T{(L)}-T{(L_j)}\right)=0
$$
which equivalently can be written as \be\label{const} \Delta
\left[\mu(L)T{(L)}\right]=0 \ee is isomorphic to the space of
generalized functions on the absolute with the isomorphism given
by the formula \be\label{solution} \phi_{T}(\chi_L)=T{(L)} \mu(L)
\ee

Investigate for the solutions of (\ref{const}) the functional
\be\label{thefunc} {1\over n}\sum_{ab}Q_{ab}= {1\over
\mu(K)}\sum_{J\in {\cal S}\backslash {\cal S}_{\rm
min}}T{(J)}\mu^2(J)\left(1-p_J^{-1}\right) \ee and the $n\to 0$
limit.

\begin{lemma}{\sl For $T{(J)}$, satisfying (\ref{const}), the functional (\ref{thefunc}) takes the form
\be\label{thefunc1} {1\over n}\sum_{ab}Q_{ab}={1\over
\mu(K)}\left(T{(K)}\mu^2(K)-\sum_{J\in
 {\cal S}_{\rm min}}T{(J)}\mu^2({J})\right)
\ee The $n\to0$ limit takes the form \be\label{limthefunc1}
\lim_{n\to0}{1\over
n}\sum_{ab}Q_{ab}=-\lim_{K\to\infty}T{(K)}=-\lim_{K\to\infty}{1\over
\mu(K)}\int_{K}\phi_{T}(x)d\mu(x) \ee where $\phi_{T}$ is the
generalized function at the absolute, corresponding to the
solution $T{(J)}$ of equation (\ref{const}). }

\end{lemma}

\noindent{\it Proof}\qquad Consider the identity
$$
{1\over \mu(K)}\sum_{J\in {\cal S}\backslash {\cal S}_{\rm
min}}T{(J)}\mu^2(J)\left(1-p_J^{-1}\right)={1\over
\mu(K)}\sum_{J\in {\cal S}\backslash {\cal S}_{\rm
min}}\Delta\left[T{(J)}\mu^2(J)\right]
$$
which follows from (\ref{const}):
$$
\Delta\left[T{(J)}\mu^2(J)\right]=T{(J)}\mu^2(J)-\sum_{j=0}^{p_J-1}T{(J_j)}\mu^2({J_j})=
$$
$$
=T{(J)}\mu^2(J)-\mu^2({J-1})\sum_{j=0}^{p_J-1}T{(J_j)}=T{(J)}\mu^2(J)-p_J^{-2}\mu^2({J})p_JT{(J)}=
$$
$$
=T{(J)}\mu^2(J)\left(1-p_J^{-1}\right)
$$
Applying the tree Newton--Leibnitz rule, we get for the functional
(\ref{thefunc})
$$
{1\over \mu(K)}\left(T{(K)}\mu^2(K)-\sum_{J\in
 {\cal S}_{\rm min}}T{(J)}\mu^2({J})\right)
$$

Compute the $n\to 0$ limit for the functional (\ref{thefunc1}).
Application of the map $\rho$ gives
$$
{1\over \mu(K)}\left(T{(K)}\mu(K)\rho(\mu(K))-\sum_{J\in
 {\cal S}_{\rm min}}T{(J)}\mu({J})\rho(\mu(J))\right)
$$

Since for $J\in {\cal S}_{\rm min}$ we have $\rho(\mu(J))=1$ up to
infinitesimal values, the non--additive expression $\sum_{J\in
 {\cal S}_{\rm min}}T{(J)}\mu^2({J})$ becomes the additive expression:
$$
\sum_{J\in {\cal S}_{\rm min}}T{(J)}\mu({J})=T{(K)}\mu(K)
$$
This implies for the functional (\ref{thefunc1})
$$
T{(K)}(\rho(\mu(K))-1)
$$
which in the $n\to 0$ limit, when $\rho(\mu(K))\to 0$ with
$K\to\infty$, takes the form
$$
-\lim_{K\to\infty}T{(K)}=-\lim_{K\to\infty}{1\over
\mu(K)}\int_{K}\phi_{T}(x)d\mu(x)
$$
where $\phi_{T}$ is the generalized function at the absolute,
corresponding to the solution $T{(J)}$ of equation (\ref{const}).
This finishes the proof of the lemma.

\bigskip

Therefore, in the frameworks of the general replica symmetry
anzats under consideration, functionals of replica matrices in the
$n\to0$ limit may take the form of the integrals over the interval
$[0,1]$ (as for the Parisi anzats), as well as the integrals of
generalized functions over ultrametric spaces.

\section{Replica symmetry breaking equation}

In the present section we, using variational procedure and the
introduced analysis on trees, find the equation, which describes
replica symmetry breaking for the Sherrington--Kirkpatrick model
in the vicinity of phase transition (in other words, when matrix
elements of the replica matrix can be considered as small
parameters). In this vicinity free energy can be decomposed into
the series of the functionals (of the type of traces of the
degrees) of the replica matrix. Functionals of this kind
(corresponding to the first several terms of the series) we
computed in \cite{RSBI}. Let us compute variations of these
functionals with respect to variations of the matrix elements
$T{(J)}$.

\begin{lemma}\label{DtraceRMst}{\sl
Variations of the following functionals have the form
\be\label{varQ2} \delta\,{\rm tr}\,Q^2=\delta\, \sum_{J\in {\cal
S}\backslash {\cal S}_{\rm
min}}{T{(J)}}^2\mu^2(J)\left(1-p_J^{-1}\right)=
$$
$$
= \sum_{J\in {\cal S}\backslash {\cal S}_{\rm
min}}\mu^2(J)\left(1-p_J^{-1}\right)2T{(J)}\delta T{(J)} \ee
\be\label{varQ21} \delta\sum_{ij}Q^4_{ij}=\delta\sum_{J\in {\cal
S}\backslash {\cal S}_{\rm
min}}{T{(J)}}^4\mu^2(J)\left(1-p_J^{-1}\right)=
$$
$$
=\sum_{J\in {\cal S}\backslash {\cal S}_{\rm
min}}\mu^2(J)\left(1-p_J^{-1}\right)4{T{(J)}}^3\delta T{(J)} \ee }
\end{lemma}

\begin{lemma}\label{DtrQ3}{\sl Variation of the cubic functional takes the form
\be\label{ultra} \delta\,\,{\rm tr}\,Q^3
=\delta\,\biggl[\sum_{L\in {\cal S}\backslash {\cal S}_{\rm
min}}\mu^3(L)\left(1-p_L^{-1}\right)\left(1-2p^{-1}_{L}\right){T{(L)}}^3+
$$
$$
+3\sum_{L\in {\cal S}\backslash {\cal S}_{\rm
min}}\mu(L)\left(1-p_L^{-1}\right) \sum_{B\in {\cal S}\backslash
{\cal S}_{\rm min}
:B<L}\mu^2(B)\left(1-p_B^{-1}\right){T{(L)}}^2T{(B)}\biggr]=
$$ $$
=\sum_{L\in {\cal S}\backslash {\cal S}_{\rm min}}\delta
T{(L)}\mu(L)\left(1-p_L^{-1}\right)\biggl[\mu^2(L)\left(1-2p^{-1}_{L}\right)3{T{(L)}}^2+
$$
$$
+ 6T{(L)}\sum_{B\in {\cal S}\backslash {\cal S}_{\rm min}
:B<L}\mu^2(B)\left(1-p_B^{-1}\right)T{(B)}+ 3\mu(L)\sum_{B:L<B\le
K}\mu(B)\left(1-p_B^{-1}\right) {T{(B)}}^2\biggr] \ee }
\end{lemma}

\noindent{\it Proof}
$$
\delta\,{\rm tr}\,Q^3 =\sum_{L\in {\cal S}\backslash {\cal S}_{\rm
min}}\mu^3(L)\left(1-p_L^{-1}\right)\left(1-2p^{-1}_{L}\right)3{T{(L)}}^2\delta
T{(L)}+
$$
$$
+3\sum_{L\in {\cal S}\backslash {\cal S}_{\rm
min}}\mu(L)\left(1-p_L^{-1}\right) \sum_{B\in {\cal S}\backslash
{\cal S}_{\rm min}
:B<L}\mu^2(B)\left(1-p_B^{-1}\right)\left(2T{(L)}T{(B)}\delta
T{(L)}+{T{(L)}}^2\delta T{(B)}\right)=
$$
$$
=\sum_{L\in {\cal S}\backslash {\cal S}_{\rm
min}}\mu^3(L)\left(1-p_L^{-1}\right)\left(1-2p^{-1}_{L}\right)3{T{(L)}}^2\delta
T{(L)}+
$$
$$
+3\sum_{L\in {\cal S}\backslash {\cal S}_{\rm
min}}\mu(L)\left(1-p_L^{-1}\right) 2T{(L)}\delta T{(L)}\sum_{B\in
{\cal S}\backslash {\cal S}_{\rm min}
:B<L}\mu^2(B)\left(1-p_B^{-1}\right)T{(B)}+
$$
$$
+3\sum_{L\in {\cal S}\backslash {\cal S}_{\rm
min}}\mu^2(L)\left(1-p_L^{-1}\right)\delta T{(L)}\sum_{B:L<B\le
K}\mu(B)\left(1-p_B^{-1}\right) {T{(B)}}^2=
$$
$$
=\sum_{L\in {\cal S}\backslash {\cal S}_{\rm min}}\delta
T{(L)}\mu(L)\left(1-p_L^{-1}\right)\biggl[\mu^2(L)\left(1-2p^{-1}_{L}\right)3{T{(L)}}^2+
$$
$$
+ 6T{(L)}\sum_{B\in {\cal S}\backslash {\cal S}_{\rm min}
:B<L}\mu^2(B)\left(1-p_B^{-1}\right)T{(B)}+ 3\mu(L)\sum_{B:L<B\le
K}\mu(B)\left(1-p_B^{-1}\right) {T{(B)}}^2\biggr]
$$
Here we used the transformation from lemma  \ref{change} (change
of order of integration in the tree integral). This finishes the
proof of the lemma.

\bigskip

Consider the functional which approximates free energy of the
Sherrington--Kirkpatrick model near phase transition, when the
replica matrix can be considered as a small parameter. This
functional, which can be obtained by decomposition of the free
energy into the Taylor series and summation over the spin degrees
of freedom, has the form \cite{MPV} \be\label{theF} F=a_2{\rm
tr}\,Q^2+a_3{\rm tr}\,Q^3+a_4\sum_{ij}Q^4_{ij} \ee where $a_2$,
$a_3$, $a_4$ are some constants. To obtain the replica solution,
one has to vary this functional, in the framework of the replica
anzats under consideration, over the parameters of the anzats and
consider the equation
$$
\delta F=0
$$
which is called the replica symmetry breaking equation.

For the replica anzats under consideration we vary the free energy
over the parameters $T{(L)}$. Combining the lemmas
\ref{DtraceRMst}, \ref{DtrQ3}, we get the following theorem:

\begin{theorem}\label{varequ}{\sl Replica symmetry breaking equation $\delta F=0$
for the free energy (\ref{theF}) in the frameworks of replica
anzats (\ref{Qmatrix}) takes the form \be\label{var_problem}
2a_2\mu(L)T{(L)}+4a_4\mu(L){T{(L)}}^3
+a_3\biggl[3\mu^2(L)\left(1-2p^{-1}_{L}\right){T{(L)}}^2+
$$
$$
+ 6T{(L)}\sum_{B\in {\cal S}\backslash {\cal S}_{\rm min}
:B<L}\mu^2(B)\left(1-p_B^{-1}\right)T{(B)}+ 3\mu(L)\sum_{B:L<B\le
K}\mu(B)\left(1-p_B^{-1}\right) {T{(B)}}^2\biggr] =0 \ee }
\end{theorem}

Equation (\ref{var_problem}) has quite complicated form. We
perform some transformations of equation (\ref{var_problem}) in
order to simplify it and find some particular solutions. Not all
solutions of the obtained new equations will be solutions of
(\ref{var_problem}) (since the performed transformations can
create additional solutions), but the correctness of the obtained
solutions may be checked separately.

Denote the LHS (left hand side) of equation (\ref{var_problem}) as
$G(L)$ and consider for this value the difference between the
value at $L$ and values at $L_j$, $j=0,\dots,p_L-1$, $L_j<L$,
$|LL_j|=1$, i.e. take the tree derivative of (\ref{var_problem}):
$$ \Delta G(L)=0
$$

\begin{lemma}\label{varequ1}
{\sl Equation $\Delta G(L)=0$ takes the form
\be\label{var_problem0}
\sum_{j=0}^{p_L-1}\left(T{(L)}-T{(L_j)}\right)\biggl\{2a_2\mu({L-1})+
4a_4\mu({L-1})\left({T{(L)}}^2+T{(L)}T{(L_j)}+{T{(L_j)}}^2\right)+
$$
$$
+3a_3\left[ -\mu^2({L-1})\left({T{(L)}}+{T{(L_j)}}\right) +  2
\sum_{B\in {\cal S}\backslash {\cal S}_{\rm min} :B\le
L_j}\mu^2(B)\left(1-p_B^{-1}\right)T{(B)}\right]\biggr\}=0 \ee }
\end{lemma}

Note that the space of solutions of (\ref{var_problem0}) contains
the space of solutions (\ref{var_problem}).

\bigskip

\noindent{\it Proof}\qquad  Easy to see that $\Delta G(L)$ has the
form \be\label{var_problem1}
2a_2\left[\mu(L)T{(L)}-\mu({L-1})\sum_{j=0}^{p_L-1}T{(L_j)}\right]+
4a_4\left[\mu(L){T{(L)}}^3-\mu({L-1})\sum_{j=0}^{p_L-1}{T{(L_j)}}^3\right]+
$$
$$
+a_3\biggl[3\left[\mu^2(L)\left(1-2p^{-1}_{L}\right){T{(L)}}^2
-\mu^2({L-1})\sum_{j=0}^{p_L-1}\left(1-2p^{-1}_{L_j}\right){T{(L_j)}}^2\right]+
$$
$$
+ 6\left[T{(L)}\sum_{B\in {\cal S}\backslash {\cal S}_{\rm min}
:B<L}\mu^2(B)\left(1-p_B^{-1}\right)T{(B)}-\sum_{j=0}^{p_L-1}T{(L_j)}\sum_{B\in
{\cal S}\backslash {\cal S}_{\rm min}
:B<L_j}\mu^2(B)\left(1-p_B^{-1}\right)T{(B)}\right]+
$$
$$
+ 3\left[\mu(L)\sum_{B:L<B\le K}\mu(B)\left(1-p_B^{-1}\right)
{T{(B)}}^2-\mu({L-1})\sum_{j=0}^{p_L-1}\sum_{B:L_j<B\le
K}\mu(B)\left(1-p_B^{-1}\right) {T{(B)}}^2\right]\biggr]  \ee

Consider the contribution
$$
T{(L)}\sum_{B\in {\cal S}\backslash {\cal S}_{\rm min}
:B<L}\mu^2(B)\left(1-p_B^{-1}\right)T{(B)}-\sum_{j=0}^{p_L-1}T{(L_j)}\sum_{B\in
{\cal S}\backslash {\cal S}_{\rm min}
:B<L_j}\mu^2(B)\left(1-p_B^{-1}\right)T{(B)}=
$$
$$
=T{(L)}\left[\sum_{j=0}^{p_L-1}\mu^2({L_j})\left(1-p_{L_j}^{-1}\right)T{({L_j})}+\sum_{j=0}^{p_L-1}\sum_{B\in
{\cal S}\backslash {\cal S}_{\rm min}
:B<L_j}\mu^2(B)\left(1-p_B^{-1}\right)T{(B)}\right]-
$$
$$
-\sum_{j=0}^{p_L-1}T{(L_j)}\sum_{B\in {\cal S}\backslash {\cal
S}_{\rm min} :B<L_j}\mu^2(B)\left(1-p_B^{-1}\right)T{(B)}=
$$
$$
=T{(L)}\mu^2({L-1})\sum_{j=0}^{p_L-1}\left(1-p_{L_j}^{-1}\right)T{({L_j})}+
\sum_{j=0}^{p_L-1}\left(T{(L)}-T{(L_j)}\right)\sum_{B\in {\cal
S}\backslash {\cal S}_{\rm min}
:B<L_j}\mu^2(B)\left(1-p_B^{-1}\right)T{(B)};
$$
To perform the transformations we used the tree Leibnitz rule
(\ref{LeibRule}) and the formula of tree derivation of the tree
integral over the higher limit (\ref{higher}).

Analogously, applying the rule (\ref{lower}) of tree derivation of
the tree integral over the lower limit, we get
$$
\mu(L)\sum_{B:L<B\le K}\mu(B)\left(1-p_B^{-1}\right)
{T{(B)}}^2-\mu({L-1})\sum_{j=0}^{p_L-1}\sum_{B:L_j<B\le
K}\mu(B)\left(1-p_B^{-1}\right) {T{(B)}}^2=
$$
$$
=\mu(L)\sum_{B:L<B\le K}\mu(B)\left(1-p_B^{-1}\right)
{T{(B)}}^2-$$ $$- \mu({L})\left[\mu(L)\left(1-p_L^{-1}\right)
{T{(L)}}^2+\sum_{B:L<B\le K}\mu(B)\left(1-p_B^{-1}\right)
{T{(B)}}^2\right]=
$$
$$
=-\mu^2(L)\left(1-p_L^{-1}\right) {T{(L)}}^2;
$$

This implies that
$$
3\left[\mu^2(L)\left(1-2p^{-1}_{L}\right){T{(L)}}^2
-\mu^2({L-1})\sum_{j=0}^{p_L-1}\left(1-2p^{-1}_{L_j}\right){T{(L_j)}}^2\right]+
$$
$$
+ 6\left[T{(L)}\sum_{B\in {\cal S}\backslash {\cal S}_{\rm min}
:B<L}\mu^2(B)\left(1-p_B^{-1}\right)T{(B)}-\sum_{j=0}^{p_L-1}T{(L_j)}\sum_{B\in
{\cal S}\backslash {\cal S}_{\rm min}
:B<L_j}\mu^2(B)\left(1-p_B^{-1}\right)T{(B)}\right]+
$$
$$
+ 3\left[\mu(L)\sum_{B:L<B\le K}\mu(B)\left(1-p_B^{-1}\right)
{T{(B)}}^2-\mu({L-1})\sum_{j=0}^{p_L-1}\sum_{B:L_j<B\le
K}\mu(B)\left(1-p_B^{-1}\right) {T{(B)}}^2\right]=
$$
$$
=3\left[\mu^2(L)\left(1-2p^{-1}_{L}\right){T{(L)}}^2
-\mu^2({L-1})\sum_{j=0}^{p_L-1}\left(1-2p^{-1}_{L_j}\right){T{(L_j)}}^2\right]+
$$
$$
+
6\biggl[T{(L)}\mu^2({L-1})\sum_{j=0}^{p_L-1}\left(1-p_{L_j}^{-1}\right)T{({L_j})}+
$$
$$
+\sum_{j=0}^{p_L-1}\left(T{(L)}-T{(L_j)}\right)\sum_{B\in {\cal
S}\backslash {\cal S}_{\rm min}
:B<L_j}\mu^2(B)\left(1-p_B^{-1}\right)T{(B)}\biggr]+
$$
$$
+ 3\left[-\mu^2(L)\left(1-p_L^{-1}\right) {T{(L)}}^2\right]=
$$
$$
=3\left[-\mu^2(L)p^{-1}_{L}{T{(L)}}^2+\mu^2({L-1})\sum_{j=0}^{p_L-1}{T{(L_j)}}^2
-2\mu^2({L-1})\sum_{j=0}^{p_L-1}\left(1-p^{-1}_{L_j}\right){T{(L_j)}}^2\right]+
$$
$$
+
6\biggl[\mu^2({L-1})\sum_{j=0}^{p_L-1}\left(1-p_{L_j}^{-1}\right)T{(L)}T{({L_j})}+
$$
$$
+\sum_{j=0}^{p_L-1}\left(T{(L)}-T{(L_j)}\right)\sum_{B\in {\cal
S}\backslash {\cal S}_{\rm min}
:B<L_j}\mu^2(B)\left(1-p_B^{-1}\right)T{(B)}\biggr]=
$$
$$
=3\left[-\mu^2({L-1})\sum_{j=0}^{p_L-1}\left({T{(L)}}^2-{T{(L_j)}}^2\right)
\right]+
6\biggl[\mu^2({L-1})\sum_{j=0}^{p_L-1}\left(T{(L)}-T{({L_j})}\right)\left(1-p_{L_j}^{-1}\right)T{({L_j})}+
$$ $$
+ \sum_{j=0}^{p_L-1}\left(T{(L)}-T{(L_j)}\right)\sum_{B\in {\cal
S}\backslash {\cal S}_{\rm min}
:B<L_j}\mu^2(B)\left(1-p_B^{-1}\right)T{(B)}\biggr]=
$$
$$
=3\sum_{j=0}^{p_L-1}\left(T{(L)}-T{({L_j})}\right)\biggl\{
-\mu^2({L-1})\left({T{(L)}}+{T{(L_j)}}\right) +
$$
$$
+ 2\biggl[\mu^2({L-1})\left(1-p_{L_j}^{-1}\right)T{({L_j})}+
\sum_{B\in {\cal S}\backslash {\cal S}_{\rm min}
:B<L_j}\mu^2(B)\left(1-p_B^{-1}\right)T{(B)}\biggr]\biggr\}=
$$
$$
=3\sum_{j=0}^{p_L-1}\left(T{(L)}-T{({L_j})}\right)\left[
-\mu^2({L-1})\left({T{(L)}}+{T{(L_j)}}\right) +  2 \sum_{B\in
{\cal S}\backslash {\cal S}_{\rm min} :B\le
L_j}\mu^2(B)\left(1-p_B^{-1}\right)T{(B)}\right].
$$

Also we get
$$
2a_2\left[\mu(L)T{(L)}-\mu({L-1})\sum_{j=0}^{p_L-1}T{(L_j)}\right]+
4a_4\left[\mu(L){T{(L)}}^3-\mu({L-1})\sum_{j=0}^{p_L-1}{T{(L_j)}}^3\right]=
$$
$$
=2a_2\mu({L-1})\sum_{j=0}^{p_L-1}\left(T{(L)}-T{(L_j)}\right)+
4a_4\mu({L-1})\sum_{j=0}^{p_L-1}\left({T{(L)}}^3-{T{(L_j)}}^3\right)=
$$
$$
=\sum_{j=0}^{p_L-1}\left(T{(L)}-T{(L_j)}\right)\left[2a_2\mu({L-1})+
4a_4\mu({L-1})\left({T{(L)}}^2+T{(L)}T{(L_j)}+{T{(L_j)}}^2\right)\right];
$$

Combining the obtained contributions for (\ref{var_problem1}), we
get for $\Delta G(L)$
$$
\sum_{j=0}^{p_L-1}\left(T{(L)}-T{(L_j)}\right)\biggl\{2a_2\mu({L-1})+
4a_4\mu({L-1})\left({T{(L)}}^2+T{(L)}T{(L_j)}+{T{(L_j)}}^2\right)+
$$
$$
+3a_3\left[ -\mu^2({L-1})\left({T{(L)}}+{T{(L_j)}}\right) +  2
\sum_{B\in {\cal S}\backslash {\cal S}_{\rm min} :B\le
L_j}\mu^2(B)\left(1-p_B^{-1}\right)T{(B)}\right]\biggr\}
$$
This finishes the proof of the lemma.

\bigskip

Equation (\ref{var_problem0}) has two families of solutions: the
first consists of the unique solution \be\label{RSym}
T{(L)}=T{(L_j)},\qquad \forall L \ee which implies that
$T{(L)}={\rm const}$. This solution (which we call the constant
solution) is the analogue, in the framework of the replica anzats
under consideration, of the known replica symmetric solution. We
will discuss solution (\ref{RSym}) in details in the next section.

The second family is related to the solutions of the equation
\be\label{new_solution} 2a_2\mu({L})+
4a_4\mu({L})\left({T{(L+1)}}^2+T{(L+1)}T{(L)}+{T{(L)}}^2\right)+
$$
$$
+3a_3\left[ -\mu^2({L})\left({T{(L+1)}}+{T{(L)}}\right) +  2
\sum_{B\in {\cal S}\backslash {\cal S}_{\rm min} :B\le
L}\mu^2(B)\left(1-p_B^{-1}\right)T{(B)}\right]=0 \ee where we made
the substitution $L_j\mapsto L$, $L\mapsto L+1$.

We consider the LHS of this equation as the function of $L$, which
we denote by $H(L)$. Consider equation, obtained by tree
derivation of (\ref{new_solution}):
$$
H(L)-\sum_{j=0}^{p_L-1}H(L_j)=0
$$
We get
$$
4a_4\left[\mu({L})\left({T{(L+1)}}^2+T{(L+1)}T{(L)}\right)
-\sum_{j=0}^{p_L-1}\mu({L-1})\left(T{(L)}T{(L_j)}+{T{(L_j)}}^2\right)\right]+
$$
$$
+3a_3\left[ -\mu^2({L})\left({T{(L+1)}}+{T{(L)}}\right)
+\sum_{j=0}^{p_L-1}\mu^2({L-1})\left({T{(L)}}+{T{(L_j)}}\right) +
2 \mu^2({L})\left(1-p_{L}^{-1}\right)T{({L})}\right]=
$$
$$
=4a_4\left[\sum_{j=0}^{p_L-1}\mu({L-1})\left(T{(L+1)}-T{(L_j)}\right)\left({T{(L+1)}}+T{(L)}
+T{(L_j)}\right)\right]-
$$
$$
-3a_3\left[ \mu^2({L})\left({T{(L+1)}}-{T{(L)}}\right)
+\sum_{j=0}^{p_L-1}\mu^2({L-1})\left({T{(L)}}-{T{(L_j)}}\right)
\right]=0.
$$
We used here the formula of tree derivation of the tree integral
over the higher limit. Dividing by $\mu({L-1})$, we get
\be\label{RSBequation}
\sum_{j=0}^{p_L-1}\biggl(4a_4\left(T{(L+1)}-T{(L_j)}\right)\left({T{(L+1)}}+T{(L)}
+T{(L_j)}\right)-
$$
$$
-3a_3\left[ \mu({L})\left({T{(L+1)}}-{T{(L)}}\right)
+\mu({L-1})\left({T{(L)}}-{T{(L_j)}}\right) \right]\biggr)=0 \ee

The obtained equation, contrary to equation (\ref{var_problem}),
does not contain tree integration and thus is much easier to
investigate. Equations (\ref{var_problem}) and (\ref{RSBequation})
are not equivalent, in particular, has non--coinciding sets of
solutions. Our aim is to find particular solutions of
(\ref{var_problem}) (taking into account the $n\to 0$ limit), and
check the relation to (\ref{var_problem}). In order to do this we
will find particular solutions of (\ref{RSBequation}).

\section{The constant solution}

In the present section we check, that the mentioned above constant
solution (\ref{RSym}), for which $T{(J)}=T={\rm const}$, indeed is
a solution of (\ref{var_problem}) (in the $n\to 0$ limit). If we
substitute (\ref{RSym}) into (\ref{var_problem}), we get
\be\label{checkRSym} a_2\mu(L)2T+a_4\mu(L)4T^3
+a_3T^2\biggl[3\mu^2(L)\left(1-2p^{-1}_{L}\right)+
$$
$$
+ 6\left[\mu(L)\mu({L-1})-\sum_{B\in {\cal S}_{\rm
min}:B<L}\mu^2({B})\right]+ 3\mu(L)(\mu(K)-\mu(L)) \biggr] =0 \ee
Here we use the following variant of the tree Newton--Leibnitz
formula:
$$
\sum_{B\in {\cal S}\backslash {\cal S}_{\rm min} :B\le
L}\mu^2(B)\left(1-p_B^{-1}\right)=\mu^2(L)-\sum_{B\in
 {\cal S}_{\rm min}:B<L}\mu^2({B})
$$

Apply to (\ref{checkRSym}) the $n\to 0$ limit (i.e. the map
$\rho$). We get
$$
a_2\mu(L)2T+a_4\mu(L)4T^3
+a_3T^2\biggl[3\mu(L)\left(\rho(\mu(L))-2\rho(\mu({L-1}))\right)+
$$
$$
+ 6\left[\mu(L)\rho(\mu({L-1}))-\sum_{B\in {\cal S}_{\rm
min}:B<L}\mu({B})\rho(\mu({B}))\right]+
3\mu(L)(\rho(\mu(K))-\rho(\mu(L))) \biggr]
$$
Taking into account that, in the $n\to 0$ limit we have
$\rho(\mu(L))\to 1$ for $L\in {\cal S }_{\rm min}$ and
$\rho(\mu(K))\to 0$ with $K\to\infty$, the expression above takes
the form (dividing by $\mu(L)$):
$$
2a_2 T+4a_4 T^3
+a_3T^2\biggl[3\left(\rho(\mu(L))-2\rho(\mu({L-1}))\right)+
$$
$$
+ 6\left[\rho(\mu({L-1}))-{1\over\mu(L)}\sum_{B\in {\cal S}_{\rm
min}:B<L}\mu({B})\rho(\mu({B}))\right]+
3(\rho(\mu(K))-\rho(\mu(L))) \biggr] =
$$
$$
=2a_2 T+4a_4 T^3 -6 a_3T^2
$$
This implies the equation
$$
2a_2T+4a_4T^3 -6a_3T^2=0
$$
which has the solution $T=0$ (trivial), and the solutions
$$
T={3a_3\pm\sqrt{9a^2_3-8a_2a_4}\over 4a_4}=\mp{a_2\over 3a_3}
$$
The last equality holds if $a_2$ is a small parameter (which is
satisfied for the Sherrington--Kirkpatrick model in the considered
regime).

Compute for the constant solution the functional
$$
{1\over n}\sum_{ab}Q_{ab}= {1\over \mu(K)}\sum_{J\in {\cal
S}\backslash {\cal S}_{\rm
min}}{T{(J)}}\mu^2(J)\left(1-p_J^{-1}\right)
$$
We get (before the $n\to 0$ limit):
$$
T{1\over \mu(K)}\sum_{J\in {\cal S}\backslash {\cal S}_{\rm
min}}\mu^2(J)\left(1-p_J^{-1}\right)=T{1\over
\mu(K)}\left(\mu^2(K)-\sum_{J\in
 {\cal S}_{\rm min}}\mu^2({J})\right)
$$
After the $n\to 0$ limit the obtained equation takes the form
$$
T{1\over \mu(K)}\left(\mu(K)\rho(\mu(K))-\sum_{J\in
 {\cal S}_{\rm min}}\mu({J})\rho(\mu({J}))\right)=$$ $$
 =T{\mu(K)\rho(\mu(K))-\mu(K)\over
 \mu(K)}\to -T
$$

\section{Solution with broken replica symmetry}

Application of the map $\rho$ to equation (\ref{RSBequation})
gives
$$
\sum_{j=0}^{p_L-1}\biggl(4a_4\left(T{(L+1)}-T{(L_j)}\right)\left({T{(L+1)}}+T{(L)}
+T{(L_j)}\right)-
$$
$$
-3a_3\left[ \rho(\mu({L}))\left({T{(L+1)}}-{T{(L)}}\right)
+\rho(\mu({L-1}))\left({T{(L)}}-{T{(L_j)}}\right) \right]\biggr)=0
$$
Since $\rho(\mu({L-1}))-\rho(\mu({L}))$ is infinitesimal, then,
omitting the infinitesimal contribution, we can put the equation
above into the form
$$
\sum_{j=0}^{p_L-1}\left(T{(L+1)}-T{(L_j)}\right)\left[4a_4\left({T{(L+1)}}+T{(L)}
+T{(L_j)}\right)- 3a_3\rho(\mu({L})) \right]=0
$$
If we assume that $T{(L+1)}$, $T{(L)}$, $T{(L_j)}$ differ
infinitesimally, then the equation above takes the form
\be\label{finequ} \left[4a_4 T{(L)}- a_3\rho(\mu({L}))
\right]\sum_{j=0}^{p_L-1}\left(T{(L+1)}-T{(L_j)}\right)=0 \ee

This equations has the following solution: \be\label{extP} T{(L)}=
{a_3\over 4a_4}\rho(\mu({L})) \ee With the standard choice of the
coefficients \cite{MPV} we get
$$
T{(L)}= {1\over 3}\rho(\mu({L}))
$$
This solution is related to the direct generalization of the
Parisi solution with broken replica symmetry onto the case of
ultrametric spaces of general form.

Remind \cite{MPV} that the Parisi solution is defined with the
help of the function on the interval $[0,1]$ of the form
$$
q(x)={1\over 3}x,\quad 0\le x\le 3T,\qquad q(x)=T,\quad 3T\le x\le
1,\qquad 0<T<<1.
$$
Actually $T$ is the constant computed in the previous Section.

We have the following theorem.

\begin{theorem}
{\sl Generalization of the Parisi solution onto the case of
general ultrametric spaces is defined as
$$
T{(L)}= {\rm min}\,\left[{a_3\over 4a_4}\rho(\mu({L})),T\right]
$$
}
\end{theorem}

\section{Appendix: trees, ultrametric spaces}

In the present section we, following papers \cite{ACHA},
\cite{Izv}, define a family of ultrametric spaces related to
trees. An ultrametric space is a metric space with the metric
$|xy|$ (the distance between $x$ and $y$), which satisfies the
strong triangle inequality
$$
|ab|\le\hbox{ max }(|ac|,|cd|),\qquad \forall c
$$

We consider directed trees, i.e. trees with partial order, which
is a direction. A partially ordered set is called {\it directed}
(and the corresponding partial order --- a direction), if an
arbitrary finite subset has the unique supremum (remind that the
supremum of the subset of a partially ordered set is a minimal
element of the set, which is larger or equal to all elements of
the subset).

Consider an arbitrary tree ${\cal T}$ (finite or infinite), such
that the path in the tree between arbitrary two vertices is
finite, and the number of edges incident to each of the vertices
is finite. If a non--maximal vertex $I\in {\cal T}$ is incident to
$p_I+1$ edges, we will say that the branching index of $I$ is
$p_I$. If maximal index $I\in {\cal T}$ is incident to $p_I$
edges, we will say that the branching index of $I$ is $p_I$.
Equivalently, branching index of a vertex $I$ in directed tree is
the number of maximal elements, which less than $I$.

The absolute of a tree will be an ultrametric space (with respect
to the naturally defined metric). Consider two equivalent
definitions of the absolute of the tree.

The first definition is as follows. The infinitely continued path
with the beginning in vertex $I$ is a path with the beginning in
$I$, which is not a subset of a larger path with the beginning in
$I$. The space of infinitely continued paths in the directed tree
${\cal T}$, which begin in some vertex $R$ (that is, the root) is
called the absolute of the tree. Obviously the definition of the
absolute of the tree does not depend on the choice of $R$ (taking
any other vertex $A$ leads to an equivalent definition).

The equivalent definition of the absolute is as follows: the
absolute is the space of equivalence classes of infinitely
continued paths in the tree ${\cal T}$, such that any two paths in
one equivalence class coincide starting from some vertex (i.e. the
tails of the paths in one equivalence class are the same). If we
choose in each of the equivalence classes the paths, which begin
in vertex $R$, we will reproduce the first definition.

We consider trees with a partial order, where the partial order is
defined in the following way. Fix the vertex $R$ and the point
$\infty$ at the absolute. To fix the point $\infty$ at the
absolute means to fix the infinitely continued path $R\infty$ from
the vertex $R$ to $\infty$. The point $\infty$ we will call the
infinite point, or the infinity. We define the following natural
partial order on the set of vertices of the tree: $J>I$ if $J$
belongs to the path $I\infty$.

Consider the absolute with excluded infinite point, or
equivalently, the space of equivalence classes of decreasing paths
in ${\cal T}$. In the following we will call the absolute with
excluded infinite point the absolute. We denote the absolute of
the tree ${\cal T}$ by $X=X({\cal T})$ (note that we already
excluded the infinite point). Let us construct the ultrametric and
the measure on $X$.

For the points $x$, $y$ of the absolute there exists a unique path
$xy$ in the tree. The notation $xy$ should be understood in the
following way. Since the points $x$, $y$ of the absolute are
identified with the paths $Rx$ and $Ry$, the path $xy$ will be
contained in $Rx \bigcup Ry$. Then there exists a unique vertex
$A$ satisfying \be\label{A} Rx=RAx,\qquad Ry=RAy,\qquad Ax\bigcap
Ay=A \ee The notation $ABC$ means that $AC=AB\bigcup BC$. Then
$$
xy=Ax\bigcup Ay
$$
Define the vertex in ${\cal T}$, which is the supremum of $x$ and
$y$: \be\label{I} {\rm sup}(x,y)=xy\bigcap x\infty \bigcap y\infty
\ee Analogously, for vertices $A$, $B$ of the tree we define
\be\label{II} {\rm sup}(A,B)=AB\bigcap A\infty \bigcap B\infty \ee
as well as for $A\in{\cal T}$, $x\in X({\cal T})$
\be\label{III}{\rm sup}(A,x)=Ax\bigcap A\infty \bigcap x\infty\ee

A partially ordered set is called {\it directed} (and the
corresponding partial order --- a direction), if an arbitrary
finite subset has the unique supremum (remind that the supremum of
the subset of a partially ordered set is a minimal element of the
set, which is larger or equal to all the elements of the subset).
Definitions (\ref{I}), (\ref{II}), (\ref{III}) make ${\cal T}$ and
${\cal T}\bigcup X({\cal T})$ the directed sets.

Put into correspondence to an edge in the tree the branching index
of the largest vertex of the edge (this definition is correct,
since any two vertices, connected by edge, are comparable). Then
the distance $|xy|$ is introduced as the product of branching
indices of edges in the directed path $RI$, $I={\rm sup}(x,y)$ in
the degrees $\pm 1$, where branching indices of increasing edges
are taken in the degree $+1$, and branching indices of decreasing
edges are taken in the degree $-1$. Here an edge is called
increasing, if the end of the edge is larger than the beginning,
and is called decreasing in the opposite case:
\be\label{distance3} |xy|=\prod_{j=0}^{N-1}
p^{\varepsilon_{I_{j}I_{j+1}}}_{I_{j}I_{j+1}},\qquad
I_0=R,\dots,I_{N}=I \ee where $\varepsilon_{I_{j}I_{j+1}}=1$ for
$I_{j}<I_{j+1}$, and $\varepsilon_{I_{j}I_{j+1}}=-1$ for
$I_{j}>I_{j+1}$.

\begin{lemma}\label{isultrametric}{\sl
The function $|xy|$ is an ultrametric (i.e. it is nonnegative,
equal to zero only for $x=y$, symmetric, and satisfies the strong
triangle inequality):
$$
|xy|\le \,{\rm max }\,(|xz|,|yz|),\qquad \forall z
$$}
\end{lemma}

To define the measure $\mu$, it is enough to define this measure
on the disks $I$, where disk $I$ is the set of all the infinitely
continued paths incident to the vertex $I$ which intersect the
path $I\infty$ only at the vertex $I$. Define the diameter $d_I$
of the disk as the supremum of the distance $|xy|$ between the
paths $Ix$ and $Iy$ in $I$. Then $I$ is the ball of radius $d_I$
with its center on any of $Ix\in I$.

\begin{definition}\label{measureeqradius}{\sl
The measure $\mu(I)$ of the disk $I$ is equal to the disk
diameter.}
\end{definition}

Since the disk $I$ contains $p_I$ maximal subdisks, which by
definitions of the ultrametric and the measure have the measure
$p_I^{-1}\mu(I)$, the measure $\mu$ is additive on disks. By
additivity  we can extend the measure on algebra generated by
disks ($\sigma$--additivity of the measure will follow from the
local compactness of the absolute, analogously to the case of the
Lebesgue measure). We denote $L^2(X,\mu)$ the space of the square
integrable (with respect to the defined measure) functions on the
absolute.

The following definition was given in \cite{RSBI}.

\begin{definition}\label{wave_type}
{\sl The subset ${\cal S}$ in a directed tree ${\cal T}$ (with the
partial order of the kind considered in the Appendix) is called of
the regular type, iff:

1) ${\cal S}$ is finite;

2) ${\cal S}$ is a directed subtree in ${\cal T}$ (where the
direction in ${\cal S}$ is the restriction of the direction in
${\cal T}$ onto ${\cal S}$);

3) The directed subtree ${\cal S}$ obey the following property: if
${\cal S}$ contains a vertex $I$ and a vertex $J$: $J<I$,
$|IJ|=1$, then the subtree ${\cal S}$ contains all the vertices
$L$ in ${\cal T}$: $L<I$, $|IL|=1$. }
\end{definition}

The maximal vertex in ${\cal S}$ we will denote $K$. We denote
${\cal S}_{\rm min}$ the set of minimal elements in ${\cal S}$.

In \cite{RSBI} the following family of replica matrices, related
to ultrametric pseudodifferential operators, was introduced
$$
Q_{IJ}=\sqrt{\mu(I)\mu(J)}T{(\,{\rm sup}\,(I,J))},\qquad I,J\in
{\cal S}_{\rm min}
$$
Here $T{(I)}$ is a function on the directed tree ${\cal T}\supset
{\cal S}$.

\bigskip

\centerline{\bf Acknowledgements}

\medskip

The authors would like to thank G.Parisi and I.V.Volovich  for
fruitful discussions and valuable comments.  One of the authors
(A.Kh.) would like to thank S.Albeverio for fruitful discussions
and support of $p$--adic investigations. This paper has been
partly supported by EU-Network ''Quantum Probability and
Applications''. One of the authors (S.V.Kozyrev) has been partly
supported by the CRDF (grant UM1--2421--KV--02), by The Russian
Foundation for Basic Research (project 05-01-00884-a), by the
grant of the President of Russian Federation for the support of
scientific schools NSh 1542.2003.1, by the Program of the
Department of Mathematics of Russian Academy of Science ''Modern
problems of theoretical mathematics'', by the grant of The Swedish
Royal Academy of Sciences on collaboration with scientists of
former Soviet Union, and by the grants DFG Project 436 RUS
113/809/0-1 and RFFI 05-01-04002-NNIO-a.

\end{document}